\documentclass[11pt,twoside]{article}
\usepackage{asp2014}

\aspSuppressVolSlug
\resetcounters

\newcommand{\beq}{\begin{equation}}
\newcommand{\eeq}{\end{equation}}

\def \gray {$\gamma$-ray }

\begin{document}

\title{Agilepy: A Python framework for scientific analysis of AGILE data}

\author{A.~Bulgarelli,$^1$ L.~Baroncelli,$^1$ A.~Addis,$^1$  N.~Parmiggiani,$^{1,2}$ A.~Aboudan,$^3$ A.~Di~Piano,$^1$ V.~Fioretti,$^1$ M.~Tavani,$^4$ C.~Pittori,$^{5,6}$ F.~Lucarelli,$^{5,6}$ and F.~Verrecchia$^{5,6}$}
\affil{$^1$INAF/OAS Bologna, Via P. Gobetti 93/3, 40129 Bologna, Italy. \email{andrea.bulgarelli@inaf.it}}

\affil{$^2$Universit\`{a} degli Studi di Modena e Reggio Emilia, DIEF - Via Pietro Vivarelli 10, 41125 Modena, Italy. }
\affil{$^2$CISAS G. Colombo University of Padua IT.}
\affil{$^4$INAF/IAPS Roma, via del Fosso del Cavaliere 100, I-00133 Roma, Italy.}
\affil{$^5$ASI Space Science Data Center (SSDC), Via del Politecnico snc, 00133 Roma, Italy.}
\affil{$^6$INAF/OAR, via Frascati 33, I-00078 Monte Porzio Catone (RM), Italy.}

\paperauthor{Andrea~Bulgarelli}{andrea.bulgarelli@inaf.it}{0000-0001-6347-0649}{INAF}{OAS}{Bologna}{BO}{40129}{Italy}
\paperauthor{Leonardo~Baroncelli}{leonardo.baroncelli@inaf.it}{0000-0002-9215-4992}{INAF}{OAS}{Bologna}{BO}{40129}{Italy}
\paperauthor{Antonio~Addis}{antonio.addis@inaf.it}{0000-0002-0886-8045}{INAF}{OAS}{Bologna}{BO}{40129}{Italy}
\paperauthor{Alessio~Aboudan}{alessio.aboudan@unipd.it}{0000-0002-8290-2184}{CISAS G. Colombo}{University of Padua}{Padova}{PD}{}{Italy}
\paperauthor{Nicol\`{o}~Parmiggiani}{nicolo.parmiggiani@inaf.it}{0000-0002-4535-5329}{INAF}{OAS}{Bologna}{BO}{40129}{Italy}
\paperauthor{Ambra~Di~Piano}{ambra.dipiano@inaf.it}{0000-0002-9894-7491}{INAF}{OAS}{Bologna}{BO}{40129}{Italy}
\paperauthor{Valentina~Fioretti}{valentina.fioretti@inaf.it}{0000-0002-6082-5384}{INAF}{OAS}{Bologna}{BO}{40129}{Italy}
\paperauthor{Marco~Tavani}{marco.tavani@inaf.it}{0000-0003-2893-1459}{INAF}{IAPS}{Roma}{RO}{00133}{Italy}
\paperauthor{Carlotta~Pittori}{carlotta.pittori@inaf.it}{0000-0001-6661-9779}{ASI}{SSDC}{Roma}{RO}{00133}{Italy}
\paperauthor{Fabrizio~Lucarelli}{fabrizio.lucarelli@inaf.it}{0000-0002-6311-764X}{ASI}{SSDC}{Roma}{RO}{00133}{Italy}
\paperauthor{Francesco~Verrecchia}{francesco.verrecchia@inaf.it}{0000-0003-3455-5082}{ASI}{SSDC}{Roma}{RO}{00133}{Italy}




  
\begin{abstract}

The Italian AGILE space mission, with its Gamma-Ray Imaging Detector (GRID) instrument sensitive in the 30 MeV–50 GeV \gray energy band, has been operating since 2007. Agilepy is an open-source Python package to analyse AGILE/GRID data. The package is built on top of the command-line version of the AGILE Science Tools, developed by the AGILE Team, publicly available and  released by ASI/SSDC. The primary purpose of the package is to provide an easy to use high-level interface to analyse AGILE/GRID data by simplifying the configuration of the tasks and ensuring straightforward access to the data.  The current features are the generation and display of sky maps and light curves, the access to \gray sources catalogues, the analysis to perform spectral model and position fitting, the wavelet analysis. Agilepy also includes an interface tool providing the time evolution of the
AGILE off-axis viewing angle for a chosen sky region. The Flare Advocate team also uses the tool to analyse the data during the daily monitoring of the \gray sky. Agilepy (and its dependencies) can be easily installed using Anaconda.
  
\end{abstract}

\section{The AGILE Observatory}

AGILE (Astrorivelatore Gamma ad Immagini LEggero) is an astrophysics mission  of the Italian Space Agency (ASI) operating since 2007 April  \citep{2008NIMPA.588...52T,2009A&A...502..995T}
 devoted to \gray and X-ray astrophysics. It carries two  instruments observing at hard X-rays between 18 and 60 keV (Super-AGILE , \citep{Feroci2007gk}) and in the \gray band between 30 MeV and 50 GeV (Silicon Tracker, \citep{Barbiellini2002jw,Prest2003280, Bulgarelli2010213, Cattaneo2011251}). The payload is completed by a calorimeter (MCAL) sensitive in the 0.4–100 MeV range \citep{2009NIMPA.598..470L} and an anticoincidence  (AC) system  \citep{2006NIMPA.556..228P}. The combination of the Silicon Tracker, MCAL, and AC forms the Gamma-Ray Imaging Detector (GRID).

\section{AGILE Science Tools}

The AGILE/GRID Science Tools developed by the AGILE Team are used to analyse \gray data starting from spacecraft files (called LOG), and the acquired events (EVT  files or event list). They provide a way to generate \gray counts, exposure and diffuse emission maps that are used as input for the binned maximum likelihood estimator (MLE).  The analysis depends on the isotropic and Galactic diffuse emission, the \gray photon statistics, and on the instrument response functions (IRFs). IRFs are matrices that characterise the effective area (Aeff), the point spread function (PSF), and the energy dispersion probability (EDP), that depend on the direction of the incoming \gray in instrument coordinates, its energy and on the on-ground event filter. The result of the MLE is an evaluation of the presence of one or more point-like or extended sources in the sky maps: this is the essential step for the scientific results of AGILE.  A full description and characterisation of the last release of the Science Tools is available in \cite{2019A&A...627A..13B}. Science Tools, IRFs and Galactic emission model are publicly available from the AGILE website at SSDC \citep{Pittori:2019cp} ( \url{http://agile.ssdc.asi.it}). 

Agilepy (\url{https://agilepy.readthedocs.io/}) is an open-source Python package to analyse AGILE/GRID data built on top of the command-line version of the AGILE/GRID Science Tools. Agilepy is similar to Fermipy \citep{Wood:2017ud} (\url{https://fermipy.readthedocs.io/})  and gammapy  (\url{https://docs.gammapy.org/}) tools, providing a common way to analyse \gray data.   

The main purpose of the package is to provide an easy to use high-level Python interface to analyse AGILE/GRID data by simplifying the configuration of the tasks and ensuring straightforward access to the data.  The current features are the generation and display of sky maps and light curves, the access to \gray sources catalogues, the analysis to perform spectral model and position fitting including the background evaluation, the aperture photometry analysis, and the wavelet analysis.   In addition, Agilepy provides an engineering interface to analyse the time evolution of the AGILE off-axis viewing angle for a chosen sky region, comparing them with Fermi/LAT off-axis evolution.  

Agilepy provides the last version of the available Science Tools (BUILD25), the H0025 instrument response functions (IRFs), and the latest version of the diffuse Galactic emission model.

Agilepy (and its dependencies) can be easily installed using Anaconda (\url{https://www.anaconda.com/}).

\section{Scientific analysis}
Many  \gray analyses can be done with few lines of configuration and code. For each functionality, a Jupyter notebook is provided with the Agilepy package to help the user in performing standard analysis by adapting the working examples.

To analyse AGILE/GRID data, first it is necessary to download data from SSDC archive (\url{https://www.ssdc.asi.it/mmia/index.php?mission=agilemmia}) and prepare two index files that list EVT and LOG files. The second step is to prepare a YAML configuration file or to use the utility method AGAnalysis.getConfiguration(), and then to use the AGAnalysis class to perform the data analysis starting from spacecraft file, event lists and IRFs. Finally, an MLE, an aperture photometry or a wavelet analysis can be performed.

The basic steps to generate maps are listed in the following example, where (glon, glat) is the centre of the region of interest (ROI) in Galactic coordinates:

\begin{verbatim}
from agilepy.api.AGAnalysis import AGAnalysis
# Creating a configuration file
confFilePath = "./agilepyConf.yaml"
AGAnalysis.getConfiguration(
    confFilePath = confFilePath, userName = "myname",
    sourceName = "3C454.3",
    tmin = 55510.0, tmax = 55530.0, timetype = "MJD",
    glon = 86.11, glat = -38.18,
    outputDir = "$HOME/agilepyAnalysis",
    verboselvl = 1,
    evtfile='$HOME/agilepyAnalysis/EVT.index',
    logfile='$HOME/agilepyAnalysis/LOG.index'
)
# Obtaining the AGAnalysis object
ag = AGAnalysis(confFilePath)
# Generate counts, exposure and diffuse emission model binned maps
maplistfile = ag.generateMaps()
# Display counts map
ag.displayCtsSkyMaps()
\end{verbatim}

To perform an MLE analysis, the second step is to prepare the list of sources to be analysed and fit the sky maps, either individually, or in a joint likelihood analysis.  Some basic steps are:

\begin{verbatim}
# Select all 2AGL source within 5 degree from the centre of the ROI
sources = ag.loadSourcesFromCatalog("2AGL", rangeDist = (0, 5))
# Free the flux parameter of 3C 454.3 (2AGL J2254+1609) source
ag.freeSources('name == "2AGLJ2254+1609"', "flux", True)
# Perform a mle analysis to fit the sky maps with 
# the prepared list of sources and standard background models
ag.mle()
# Print results with significance > 0
ag.selectSources("sqrtTS > 0")
\end{verbatim}

It is possible to specify how mle() method works using the \textit{mle} section of the YAML configuration file. It is also possible to evaluate and fix the isotropic and diffuse emission models. The preparation of a light curve using MLE or aperture photometry methods is straightforward. An example for the MLE method would be the following:

\begin{verbatim}
ag.lightCurveMLE("2AGLJ2254+1609", binsize=86400)
\end{verbatim}

The result is shown in Figure \ref{fig:architecture}.

\articlefigure[width=.98\textwidth]{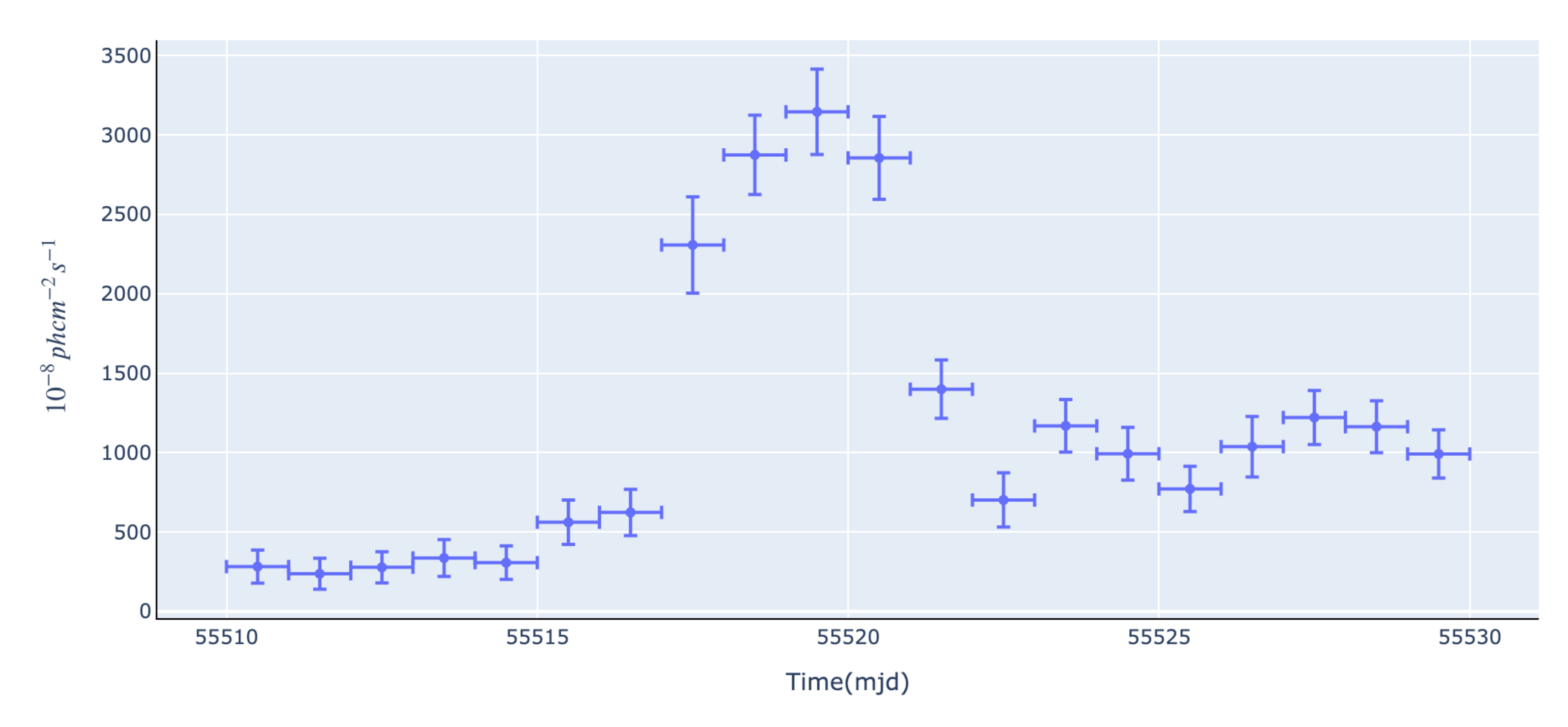}{fig:architecture}{ The light curve of 3C454.3 source generated with Agilepy. }

\section{Conclusions}
The main purpose of Agilepy is to facilitate the use of the AGILE/GRID Science Tools providing an easy to use high-level Python interface to analyse AGILE/GRID data. The Flare Advocate team uses Agilepy to analyse the data during the daily monitoring of the \gray sky.  New features are planned for future releases.

\acknowledgements The AGILE Mission is funded by the Italian Space Agency (ASI) with scientific and programmatic participation by the Italian Institute
of Astrophysics (INAF) and the Italian Institute of Nuclear
Physics (INFN). Investigation supported by the ASI grants I/089/06/2, I/028/12/5 and I/028/12/6.
We thank the ASI management for unfailing support during AGILE operations.
We acknowledge the effort of ASI and industry personnel in operating
the  ASI ground station in Malindi (Kenya), and the data processing done
at the ASI/SSDC in Rome: the success of AGILE scientific operations depends on the effectiveness of the data flow from Kenya to SSDC and INAF/OAS Bologna and the data analysis and software management.


\bibliography{P10-123}





\end{document}